\documentclass{ieeetj}
\JournalAcronym{OJSP}
\usepackage{cite}
\newif\ifreferencesused
\referencesusedfalse
\let\originalcite\cite
\renewcommand{\cite}[2][]{%
    \global\referencesusedtrue
    \if\relax\detokenize{#1}\relax
        \originalcite{#2}%
    \else
        \originalcite[#1]{#2}%
    \fi
}
\usepackage{amsmath,amssymb,amsfonts}
\usepackage{algorithmic}
\usepackage{booktabs}
\usepackage{graphicx,color}
\graphicspath{{figures/generated_v2/}{figures/external/FigureSet_v0/}{figures/external/}{figures/}}
\usepackage{textcomp}
\usepackage{xcolor}
\usepackage{hyperref}
\hypersetup{hidelinks}
\usepackage{algorithm,algorithmic}
\makeatletter
\renewcommand{\@IEEEfigurecaptionsepspace}{\vskip-5pt\relax}
\makeatother
\def\BibTeX{{\rm B\kern-.05em{\sc i\kern-.025em b}\kern-.08em
    T\kern-.1667em\lower.7ex\hbox{E}\kern-.125emX}}
\AtBeginDocument{\definecolor{NavyBlue}{RGB}{0,86,125}}

\def\OJlogo{\vspace{-4pt}$<$Society logo(s) and publication title will appear here.$>$}
\def\seclogo{\vspace{10pt}$<$Society logo(s) and publication title will appear here.$>$}
\def\authorrefmark#1{\ensuremath{^{\textbf{#1}}}}

\begin{document}
\receiveddate{XX Month, XXXX}
\reviseddate{XX Month, XXXX}
\accepteddate{XX Month, XXXX}
\publisheddate{XX Month, XXXX}
\currentdate{XX Month, XXXX}
\doiinfo{XXXX.2022.1234567}

\markboth{}{Avramidis {et al.}}

\title{Tracking Driving Stressors through Multimodal Physiological Monitoring
\thanks{
This work involved human subjects in its research. Participant consent was obtained and related research was performed in accordance with institutional guidance from Toyota Legal One.

Code is available at https://github.com/usc-sail/driving-stressors.}
}

\author{
  \IEEEauthorblockN{
    Kleanthis Avramidis\authorrefmark{1},\quad
    Emily Zhou\authorrefmark{1},\quad
    Tiantian Feng\authorrefmark{1},\quad
    Hossein Hamidi Shishavan\authorrefmark{2},\\
    Frederico Marcolino Quintao Severgnini\authorrefmark{2},\quad
    Danny J. Lohan\authorrefmark{2},\quad
    Paul Schmalenberg\authorrefmark{2},\\
    Ercan M. Dede\authorrefmark{2},\quad
    Shrikanth Narayanan\authorrefmark{1}
  }
  \affil{Signal Analysis and Interpretation Lab,
    University of Southern California, Los Angeles, USA
  }
  \affil{Toyota Research Institute of North America, Ann Arbor, USA
  }
}

\begin{abstract}
Understanding and mitigating driving stress is important for improving road safety and driver well-being. Reliable estimation, however, requires distinguishing biobehavioral responses to individual stressors from gradual physiological and contextual changes. We collected physiological data and vehicle telemetry from 31 participants across 44 simulated-driving sessions containing controlled stressor events. Under cross-validation, a multimodal classifier achieved an AUROC of 0.768 when distinguishing the stressor phase from an earlier baseline, reflecting both stressor effects and temporal drift. Controlling for drift retained an AUROC of 0.661, but revealed stronger responses to sustained than brief stressors, and shifted feature attribution toward phasic cardiac and electrodermal markers. We further quantified the interaction between model-estimated physiological stress and observable changes in vehicle control through simulation telemetry. Our findings show that stressor-aware modeling can identify physiologically grounded responses that correspond to meaningful changes in driving behavior.
\end{abstract}


\maketitle

\begin{figure}
    \centering \includegraphics[width=\linewidth]{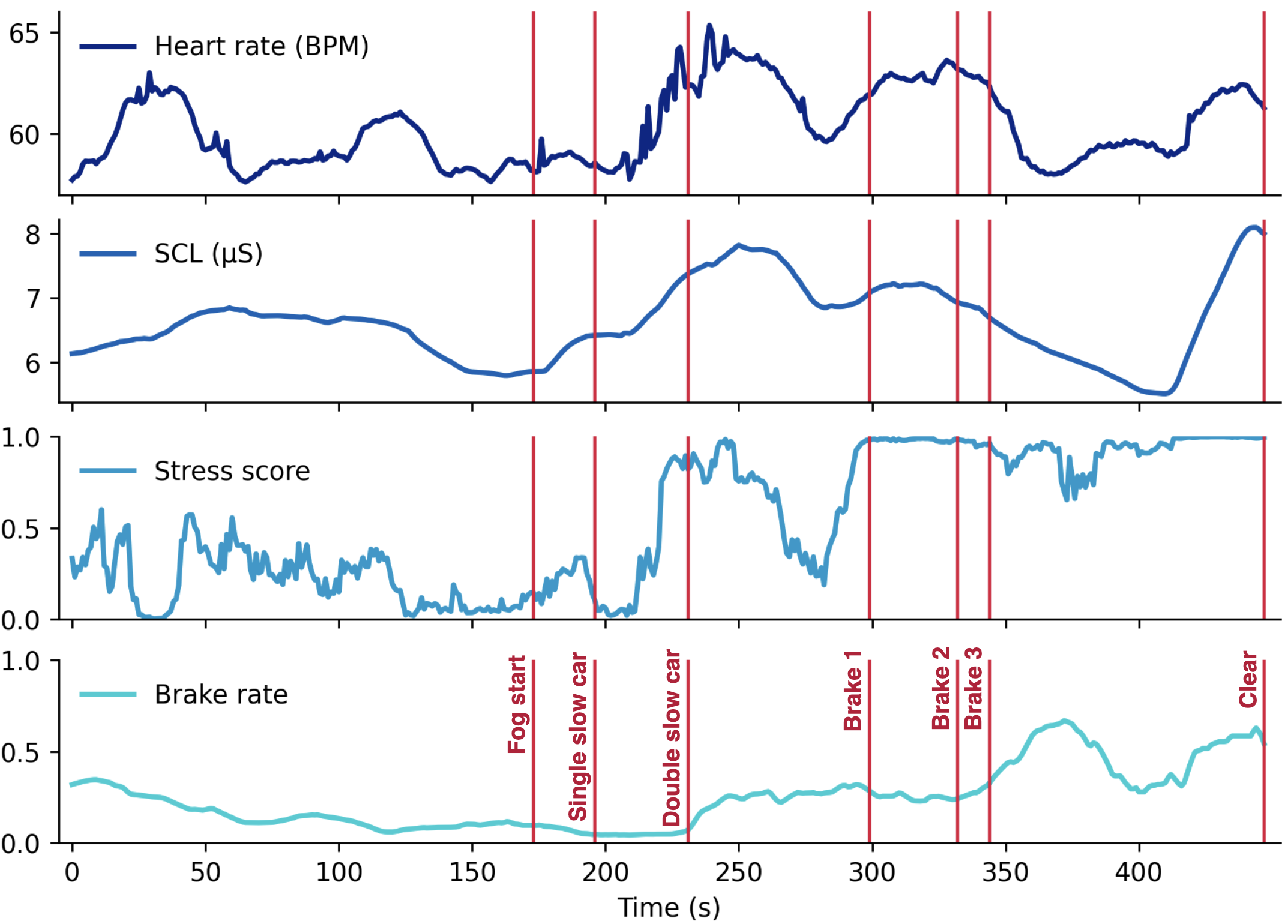}
    \caption{\textbf{Visualization of physiological signals, stress predictions, and driving behavior during a sample session from the developed driving simulation}. Vertical red lines indicate the onset of driving events, including environmental challenges (e.g., dense fog), vehicle interactions (e.g., slow car scenarios), abrupt braking events, up to the return to clear conditions. The plots illustrate the interplay between physiological responses, model-predicted stress, and behavioral adjustments across the driving context, which this study attempts to interpret.}
    \label{fig:header}
\end{figure}

\section{Introduction}

Driving safety is a critical societal concern since it affects millions of individuals daily, irrespective of being drivers or pedestrians. According to the World Health Organization~\cite{world2019global}, motor vehicle crashes are a leading cause of death in the United States, with over 120 casualties every single day. While safety is often measured in terms of crash statistics, a more pervasive but less visible factor is psychological strain. Adverse driving conditions such as congestion, time pressure, or aggressive behavior can induce significant levels of stress. Even when not resulting in accidents, stress or fatigue could trigger mental health issues, including risk of depression, and decline in quality of life~\cite{gee2004traffic,useche2018work}. The importance of safe vehicle operation is reflected in the application of technological advances aimed at mitigating risks on the road, including advanced braking systems, adaptive cruise control, automatic collision avoidance, and driver alerts. However, reducing traffic accidents maps to the inherent uncertainty in modeling and predicting human behavior~\cite{bone2017signal} and driver-vehicle interaction.

Driving demands continuous behavioral regulation shaped by both psychological states and situational dynamics. Real-world stressors, including abrupt braking, sharp maneuvers, or unexpected obstacles, directly challenge a driver’s capacity to sustain safe control of the vehicle. While modern safety frameworks frequently leverage behavioral cues, e.g., pedal dynamics, steering variability~\cite{nvemcova2020multimodal}, or gaze patterns~\cite{friedrichs2010camera}, to flag potential risk~\cite{liang2007real}, these indicators alone often suffer from signal noise and substantial inter-individual variability~\cite{engstrom2017effects}. Notably, although considerable research has addressed distraction~\cite{taamneh2017multimodal}, fatigue, and drowsiness~\cite{borghini2014measuring,sikander2018driver}, relatively fewer studies have explicitly examined stress within  driving, likely because stress responses inherently elicit rapid physical actions (e.g., emergency braking) that may confound the underlying physiological markers. In this study, we center our attention on short-term driving events that act as transient stressors and probe how autonomic nervous system activity co-evolves with behavioral adjustments. We introduce a data acquisition protocol and computational pipeline to capture and interpret these multimodal interactions (see sample visualization in Figure~\ref{fig:header}), with the goal of improving the precision and reliability of driving stress detection.

Signal processing and Artificial Intelligence (AI) techniques are essential components of computational systems for estimating psychological stress from physiological signals. AI methods are increasingly being integrated into healthcare and well-being applications to model aspects of the human condition~\cite{gershon2016daily,yuan2024self,thapa2024sleepfm}. In the context of driving, this potential for improving driver safety and support is significant. However, deploying such systems in practice remains challenging due to the difficulty of collecting data in heterogeneous driving conditions and the corresponding variability in human responses~\cite{engstrom2017effects}, which often limits model generalizability. Further, while many previous works have proposed AI-based systems for driving safety~\cite{healey2005detecting,solovey2014classifying,nvemcova2020multimodal}, few have examined the implications of how such systems are evaluated. Studies typically contrast a baseline period against subsequently induced stressors, so that any physiological quantity that drifts gradually over a session will separate those two whether or not a stressor occurred. 

\begin{figure}
    \centering
    \includegraphics[width=\linewidth]{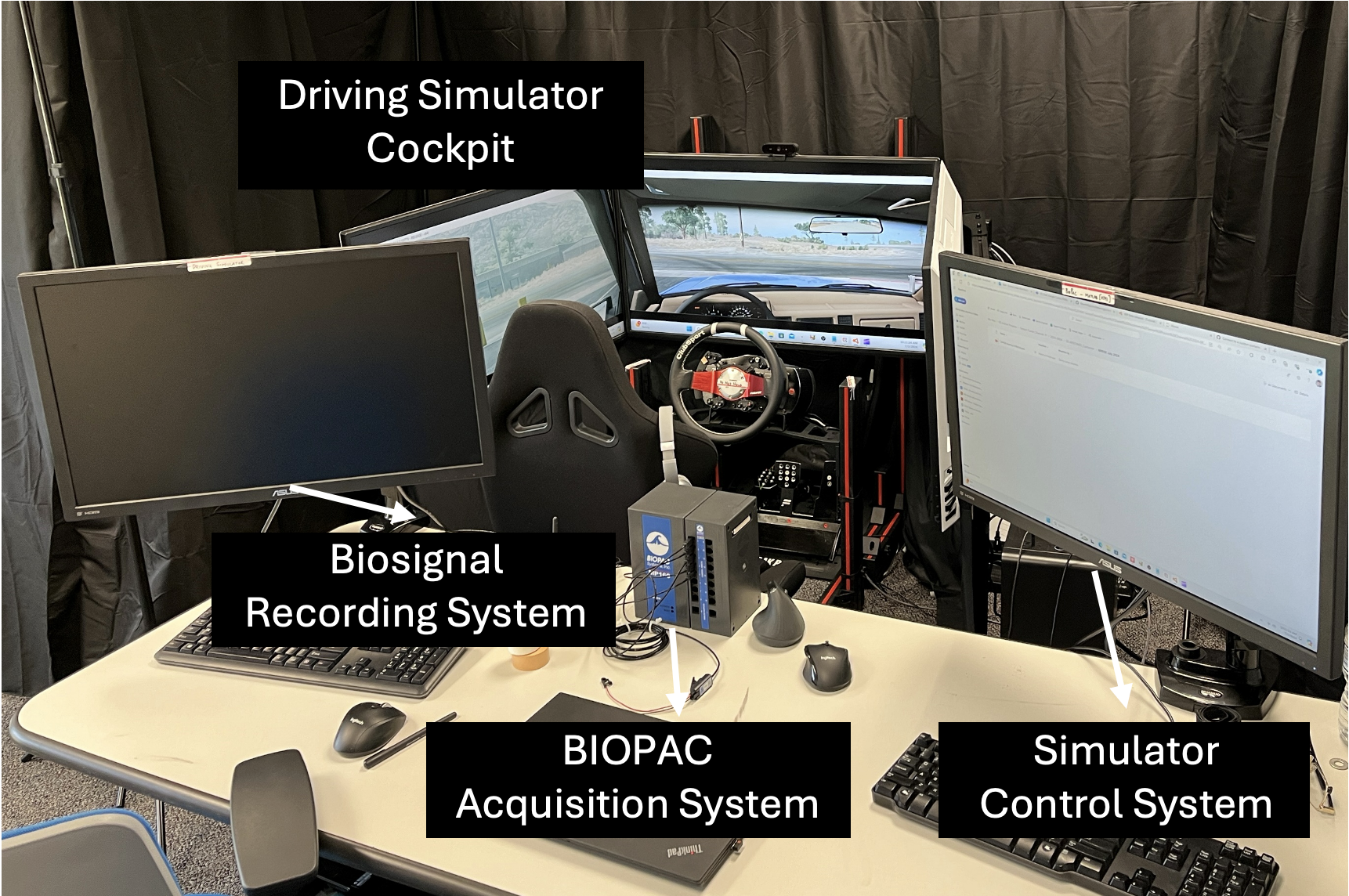} 
    \caption{\textbf{Driving simulator used in the study.} It features the cockpit with driving hardware and displays, and two systems for capturing biosignals and behavioral data, alongside the processing of the driving scenarios.}
    \vspace{-5mm}
    \label{fig:simulator}
\end{figure}

\begin{figure}
    \centering
    \includegraphics[width=\linewidth]{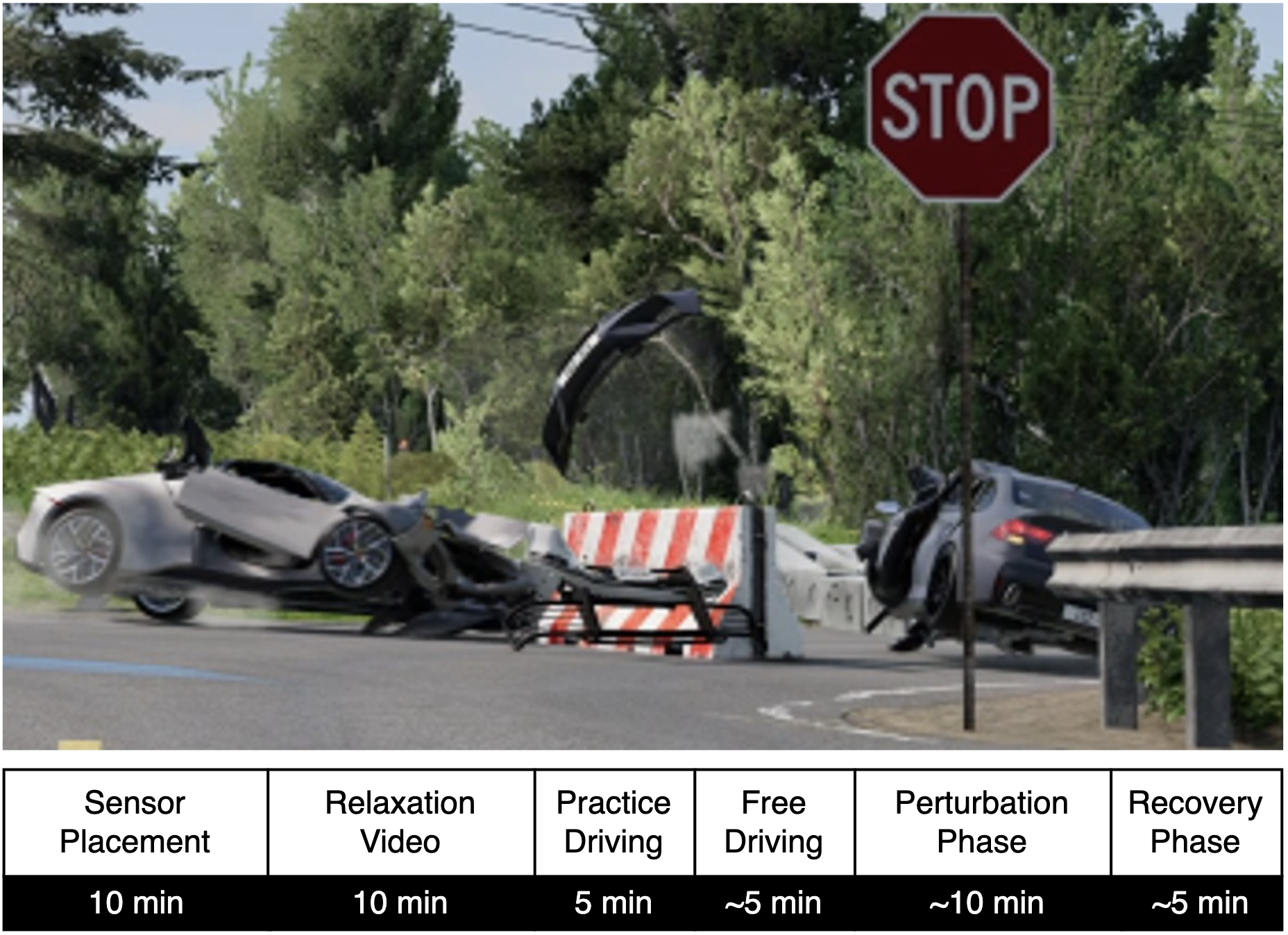} 
    \caption{\textbf{Details of the driving simulation}: Top: Example of stressor event (crash -- surprise scenario). Bottom: Timeline of the experimental procedure, including the initial setup and the three driving phases (video baseline, practice driving, and the full experiment) and their duration. Stressors are introduced throughout the perturbation phase.}
    \vspace{-2mm}
    \label{fig:sim-crash}
\end{figure}

In this study, we leverage a driving simulator (Figure~\ref{fig:simulator}) with timestamped stressor events (Figure~\ref{fig:sim-crash}) to investigate their influence on driving behavior and treat this ambiguity in model evaluation as a measurement problem. The contributions of our study are outlined below:

\begin{itemize}
    \item We present a study design protocol and collection of physiological data from 31 participants during simulated driving under induced stress conditions.
    \item We develop a multimodal machine learning framework to estimate stressor presence and quantify how much of model accuracy is attributable to temporal drift over time. The base model reaches 0.768 AUROC.
    \item We validate two control mechanisms, namely a per-recording baseline and a contrast drawn from within the stressor period, to remove the temporal effect, through which we retain 0.661 AUROC while better isolating the effects of individual stressors.
    \item We provide insight into which forms of stressor events have detectable effects, how repeated exposure alters the response, and how these are correlated with measures such as control of steering wheel and pedal usage.
    \vspace{-4mm}
\end{itemize}

\section{Background}

The detection of driving stressors through physiological monitoring has long been studied for improving road safety. This approach leverages the inherent link between a driver's physiological state and their behavioral responses that influence driving performance. The use of physiological measures of the autonomic nervous system has been extensively explored for this purpose. In particular, heart rate (HR) and heart rate variability (HRV) are well-known markers of the sympathetic and parasympathetic system interplay~\cite{shaffer2017overview}. Increased HR and decreased HRV are often associated with stress, fatigue, and cognitive load~\cite{kim2018stress}, all of which can impair driving experience~\cite{engstrom2017effects,arutyunova2024heart}. Electrocardiography (ECG) and photoplethysmography (PPG) are the gold standards for measuring heart rate activity. On the other hand, electrodermal activity (EDA) reflects changes in skin conductance due to sweat gland activity. It is considered a stable stress marker as it maps to sympathetic activations of the nervous systems. In particular, EDA peaks are typically correlated with physiological arousal~\cite{affanni2018driver, avramidis2023multimodal}, emotional responses~\cite{boucsein2012electrodermal}, and cognitive workload~\cite{visnovcova2016complexity}. EDA sensors, being minimally invasive, have been widely adopted in studies of driving behavior~\cite{choi2017wearable,rigas2008reasoning}.

While used less frequently, other sensor modalities, such as respiratory measures and skin temperature have also shown complementary evidence of physiological responses under stress. For example, increased respiratory rate and irregular breathing patterns can signify stress, anxiety, or fatigue~\cite{solaz2016drowsiness, tiwari2019breathing}. Skin temperature has also been investigated as a marker of emotional arousal. While less variable than other physiological signals, changes in skin temperature can reflect regulatory responses in sympathetic activity~\cite{ioannou2014thermal} which can either map to stress or fatigue~\cite{yamakoshi2007preliminary,or2007development}. However, several of those modalities change gradually over the course of a recording for reasons unrelated to affect. Tonic electrodermal activity is sensitive to electrode polarization and to the progressive hydration of the skin beneath the electrode~\cite{bari2026electrodermal,society2012publication}. Peripheral skin temperature is strongly influenced by ambient thermal conditions that vary over the course of a session. Sustained performance of a task also produces gradual shifts in arousal that are not tied to any particular event~\cite{warm2008vigilance}. Each of these conditions unfolds gradually over time and should be regressed out from model evaluation in relevant studies.

The integration of physiological measures holds promise for developing comprehensive systems capable of detecting driving stressors and enhancing road safety. Most studies leverage classical machine learning (ML) algorithms, such as Support Vector Machines (SVMs) and Random Forests~\cite{nvemcova2020multimodal}, to classify driver states from knowledge-driven physiological features. More recently, deep learning models have gathered significant attention in driving stressor estimation studies. Their ability to automatically extract and classify patterns from raw sequential data has improved the detection of driver fatigue and cognitive load~\cite{zeng2018eeg}. However, despite their encouraging results emerging on individual datasets, the inherent limitations of training data availability and the relatively low information density within sensor time-series data present significant challenges in reliable stressor estimation using physiological signals. Consequently, here we adopt a knowledge-driven ML framework which allows for the extraction of interpretable psychophysiological markers.

\section{Study Design}

\begin{table*}[t]
\centering
\caption{Summary of simulated driving stressors employed in the experiment. Each scenario elicited a specific stress response through variations in driving context. Compliance refers to the fraction of sessions in which the drivers met with those conditions.}
\label{tab:scenarios}
\begin{tabular}{llcc}
\toprule
\textbf{Stressor Event} & \textbf{Description} & \textbf{Scenario} & \textbf{Compliance} \\
\midrule
Timer, Pace Car & Lead vehicle enforces reduced speed under time pressure, inducing frustration and impatience. & Impatience & 30\,/\,30 \\
Delivery Van & A delivery van repeatedly stops and brakes, delaying the driver stuck behind. & Impatience & 75\,/\,90 \\
Construction & Lane closures and halted progress cause prolonged wait times, with no control for the driver. & Impatience & 30\,/\,30 \\
\midrule
Car Crash & Simulated vehicle collision with abrupt audiovisual cues. & Surprise & 15\,/\,15 \\
Barrel Explosion & AI-driven vehicle hits a barrel, triggering an explosion in the opposite lane. & Surprise & 14\,/\,15 \\
\midrule
Dense Fog & Simulated dense fog condition reduces visibility and heightens caution. & Irritation & 15\,/\,15 \\
Slow Parallel Cars & AI-driven vehicles are stuck in front of the driver, moving in slow speeds. & Irritation & 22\,/\,30 \\
Sudden Braking & AI-driven vehicle ahead performs erratic and abrupt maneuvers, including sudden braking. & Irritation & 34\,/\,75 \\
\bottomrule
\end{tabular}
\vspace{-2mm}
\end{table*}

\subsection{Participants and Protocol}

We built our stressor estimation ML model on a dataset of simulated driving, collected between 01/06/2024 and 30/11/2024 from 32 participants (9 females) in a total of 45 unique annotated sessions. Participants' age ranged from 20 to 45 years. After manual inspection, a single participant was excluded from further analysis due to compromised signal quality, leaving 44 sessions in total. All study procedures adhered to institutional guidelines and were approved by \textit{Toyota Legal One}. All participants provided informed consent before data collection. The study was conducted using a three-screen immersive driving simulator equipped with a steering wheel, acceleration and brake pedals, and an adjustable chair with headphones for audio feedback (Figure~\ref{fig:simulator}). Participants underwent a 10-minute preparation, during which they gave informed consent, were provided with procedural guidelines, and fitted with all the physiological sensors.

\subsection{Signal Acquisition}

Physiological signals were recorded using a BIOPAC system at a 2\,kHz sample rate. We collected five modalities in total: ECG, PPG, EDA, respiration and skin temperature. ECG electrodes were positioned on the chest to capture cardiac activity. PPG was used to correct heart rate from ECG and was recorded from the index fingertip. EDA was measured via electrodes on the palm, while the temperature sensor was placed on the nasal tip to track autonomic thermal responses. Respiratory belts recorded abdominal and thoracic expansion, from which we extracted their average as the respiration signal. To minimize motion artifacts, participants controlled the simulator using only their non-instrumented hand. Vehicle telemetry was logged by the simulator over the same interval, providing speed, engine speed, steering angle, and throttle and brake pedal traces at a rate of 7\,Hz.

\subsection{Driving Simulation}

The simulation was implemented using BeamNG.tech~\cite{beamng_tech}, selected for its realistic vehicle dynamics and high-fidelity modeling of the driving environment, as in Figure~\ref{fig:sim-crash}. The experiment consisted of three parts: (1) a 10-min baseline phase where participants watch a video of a car driving with calm background music, (2) a 5-minute practice session to familiarize participants with the equipment, and (3) the main experimental phase, designed to induce stress and emotional responses through diverse driving scenarios. The session included baseline reassessment (free driving), exposure to emotionally stimulating conditions, and a recovery period to observe regulation processes.

The three phases of a session occur in a fixed order that is identical for every participant: free driving, then the stressor phase, then recovery. This ordering renders the free-driving segment as a within-session baseline recorded under the same instrumentation as the stressor phase. However, because every stressor window is later in the session than every free-driving window, any quantity that varies monotonically with time in session is confounded with the stressor manipulation. Therefore, we introduce evaluation protocols that control for this effect and quantify how much of the stress effect is attributable to time. Importantly, because recovery is the latest phase in the session and contains no stressors, we use it to dissociate monotone artifacts from genuine stressor-locked responses.


\begin{table}[t]
\centering
\caption{Set of features extracted per modality using a 20-second window (10 seconds before and after each time point).}
\label{tab:physio_features}
\small
\begin{tabular}{@{}lp{6.8cm}@{}}
\toprule
\textbf{Modality} & \textbf{Extracted Features} \\
\midrule
\textbf{ECG, PPG} & Heart Rate (HR, bpm), HR Range (bpm), RMSSD (ms), Pulse Wave Amplitude (PWA, a.u.), PWA Slope (a.u./s), Pulse Transit Time (PTT, ms), median and std \\
\textbf{EDA} & SCL mean ($\mu$S), slope ($\mu$S/s), SCR Rate (peaks/min), SCR Amplitude ($\mu$S), SCR Rise Time (s), SCR Area Under Curve ($\mu$S$\cdot$s), SCR Recovery Time (s) \\
\textbf{RSP} & Respiratory Period (s), and Depth (a.u.), Period Variability (s), Inspiration:Expiration Ratio, Breath-Hold Fraction, Thoracoabdominal Phase Angle (deg) and Coherence \\
\textbf{SKT} & Mean Temperature ($^{\circ}$C) and Slope ($^{\circ}$C/s) \\
\midrule
\textbf{Vehicle} & Vehicle and Engine Standard Deviation (rpm), Steering Angle Standard Deviation (deg), Steering Usage Entropy, Throttle Rate, Throttle Magnitude, Throttle Usage Entropy, Brake Rate, Brake Magnitude, Brake Usage Entropy \\
\bottomrule
\end{tabular}
\vspace{-4mm}
\end{table}

\subsection{Elicited Stressors}

We employed three strategies to elicit emotional responses of \textbf{impatience~(M)}, \textbf{surprise~(S)}, and \textbf{irritation~(I)}. The first (M) utilized monotonous driving, where extended low-stimulation driving periods were designed to elicit impatience. This was achieved by incorporating a slow-moving delivery van and two construction zones, creating an environment that gradually induced frustration due to prolonged inactivity and lack of engagement. The second approach (S) included sudden crashes and explosions that were designed to evoke surprise, hence disrupting the drivers' expectations. The third approach (I) combined the two methods, creating a dynamic experience that included aggressive virtual drivers performing reckless maneuvers, alongside slower stressors like fog and obstructive vehicles moving at reduced speeds. This approach forced participants to navigate frustrating and restrictive conditions, thereby eliciting irritation~(I).


\section{Methods}

We hypothesized that the aforementioned elicited stressors would modulate the emotional and behavioral state of the participants, resulting in changes that would map to their recorded physiological responses. In the following, we describe the ML framework used to identify such physiological markers and differentiate them from the free-driving condition. We considered 4 sources of physiological activity, namely cardiac activity through ECG and PPG, electrodermal activity through EDA, respiration (RSP), and thermoregulation through skin temperature (SKT).

\subsection{Data Pre-processing}

Biosignals were recorded from a person's nose (SKT), chest (ECG, RSP) and instrumented hand (PPG, EDA) and were visually inspected for major artifacts. ECG recordings were detrended using a 5th-order high-pass Butterworth filter with a 0.5\,Hz cutoff, followed by powerline interference removal at 60\,Hz. EDA signals were denoised using a 4th-order low-pass Butterworth filter with a 3\,Hz cutoff. The thoracic and abdominal belt signals were averaged point-wise and the average was band-pass filtered between 0.05\,Hz and 3\,Hz using a 2nd-order Butterworth filter. No pre-processing was applied to the skin temperature signal. All filtering methods were based on the NeuroKit2 library~\cite{Makowski2021neurokit}. Finally, all signals were low-pass filtered at 100\,Hz with a 4th-order Butterworth anti-aliasing filter and then downsampled from 2000\,Hz to 250\,Hz. Vehicle telemetry was left at its native rate.

\subsection{Feature Extraction}

As shown in Table~\ref{tab:physio_features}, 23 features related to physiological stress~\cite{nvemcova2020multimodal} were extracted from the electrocardiogram (ECG), photoplethysmogram (PPG), electrodermal activity (EDA), respiratory (RSP), and skin temperature (SKT) recordings. Each feature point was computed using a 20-second signal window centered on the middle point. This window length was selected to balance temporal fidelity and feature reliability. We also used a short hop size of 1 second between windows. This not only increased the effective training set size but also enabled temporally resolved predictions.

Heart rate (HR) was estimated independently from ECG R-peaks and PPG pulse peaks and averaged within each window; when one modality was unavailable, the other was used alone. We also extracted HR range and RMSSD, pulse-wave amplitude and slope, and the median and standard deviation of pulse transit time from each ECG R-peak to the corresponding PPG foot. EDA was decomposed into tonic skin conductance level (SCL) and phasic skin conductance responses (SCRs) using cvxEDA~\cite{greco2015cvxeda}. EDA features comprised mean and slope of SCL and frequency, mean amplitude, rise time, integrated area, and recovery time of SCRs. Respiratory features included period, depth, inspiration-to-expiration ratio, period variability, breath-hold fraction, and thoracoabdominal phase angle and coherence. We used respiratory period rather than rate to avoid numerical instability in windows containing only 1--2 cycles, consistent with prior work~\cite{bach2016linear}. Finally, skin-temperature features comprised its mean and slope, reflecting peripheral vasodilation or constriction. SCL and temperature slopes were estimated by linear regression within each window.

Finally, vehicle features were computed on the simulator's native telemetry over the same windows and comprise average speed, speed standard deviation, engine speed standard deviation, steering angle standard deviation, steering usage entropy, and the rate, magnitude and usage entropy of the throttle and brake pedals. Measures that could not be estimated within a 20\,s window were imputed with a $k$-nearest-neighbor imputer fitted on the training sets. 
\vspace{-2mm}

\subsection{Model \& Training Protocol}

Our modeling objective was to use the computed features to estimate short- and long-term physiological stress likelihood associated with stressor-rich driving periods. To achieve that, we trained a gradient boosting classifier (XGBoost~\cite{chen2016xgboost}) to differentiate between stressor-rich and baseline free-driving segments. The practice driving phase was discarded. We used a maximum tree depth of 2, an L2 regularization parameter of 50, 200 estimators and a learning rate of 0.05. To further account for class imbalance during training, we calculated and applied inverse class-frequency weighting.

The classifier was trained on the 23 physiological features using a leave-one-subject-out (LOSO) cross-validation (CV). In each fold, the held-out participant was used exclusively for testing. To assess robustness to variability in training-subject composition, we further applied subject-level resampling: for each LOSO fold, $N-1$ training participants were sampled with replacement, and the model was trained on this resampled set across 10 random seeds per fold. No model hyperparameters were tuned on the evaluation metric.

\subsection{Recording Normalization}

Prior to training, all feature vectors were standardized independently for each recording, using feature-wise median and interquartile range (IQR). We considered three candidate baselines to compute these parameters: \texttt{free60}, i.e., the first minute of free driving, assimilating most prior work; \texttt{practice}, the separate practice drive recorded before the protocol; and \texttt{session}, the whole free-driving and stressor recording. Statistics are computed once per recording and applied to every segment of that recording. Three sessions were excluded from all model-based analyses because of insufficient recording length; the telemetry contrast of Section~V.\ref{sec:behavior} uses all 44 sessions. We verified that our results do not change substantially under this pre-processing choice.

\subsection{Evaluation Protocol}

The model was trained end to end and predictions for each test participant were derived per second of their recording. During inference, we measured the model's ability to identify external stressors by assigning them larger probability of representing unsafe states compared to respective control conditions. To this end, we used the area under the receiver-operating curve (AUROC) as our primary metric to differentiate the free-driving from the stressor phase. To distinguish between temporal drift and genuine stress responses, we further report two custom metrics:

\emph{Within-phase contrast} provides the drift-controlled estimate. Within the stressor phase, windows centered 10--40 seconds after an event onset are contrasted with control windows centered more than 30 seconds from every onset and outside the post-onset period of any event. In every session, the control windows lie later in the phase on average than the post-onset windows because onsets are concentrated in its earlier part. To test whether this contrast is solvable from position in the phase alone, we circularly shifted each session's onset sequence within the stressor phase, preserving onset count and spacing. No shifted-onset model reached above-chance (AUROC$\,=0.519$).

\emph{Phase drift} is the Spearman correlation between the predicted stress and time within a non-stressor phase alone (either free driving or recovery), so that any monotone trend is by construction a drift artifact.

Confidence intervals were obtained using 1,000 bootstrap iterations with participants as the resampling unit. To derive a per-participant score we simply averaged all sessions of each participant before hypothesis testing. Statistical comparisons of AUROC against chance were performed using Wilcoxon signed-rank tests over participant-level scores. When a stressor occurred multiple times within a session, scores from its repeated occurrences were first averaged within session. Whenever necessary, we controlled for multiple comparisons using the false discovery rate (FDR).

\section{Results \& Discussion}

\begin{table}[t]
\centering
\caption{Effect of the per-recording baseline, with 95\% bootstrap confidence intervals over participants. \emph{Phase} contrasts free driving against the stressor phase, and \emph{within-phase} contrasts post-onset against control windows drawn from the stressor phase alone. \emph{Drift} is the median per-session Spearman correlation between predicted stress and time within free driving, where no stressor occurs. \emph{Recovery} contrasts the stressor-free recovery phase against free driving.}
\label{tab:baselines}
\setlength{\tabcolsep}{5.5pt}
\small
\begin{tabular}{@{}lcccc@{}}
\toprule
Baseline & Phase & Within-phase & Drift $\rho$ & Recovery \\
\midrule
\texttt{free60}  & 0.768 & 0.587 & $+0.492$ & \textbf{0.803} \\
 & (0.699--0.829) & (0.520--0.643) & & (0.723--0.873) \\
\addlinespace
\texttt{practice} & 0.610 & 0.554 & $+0.001$ & 0.583 \\
 & (0.548--0.670) & (0.488--0.610) & & (0.496--0.664) \\
\addlinespace
\texttt{session} & 0.665 & \textbf{0.661} & $+0.158$ & 0.634 \\
 & (0.586--0.743) & (0.594--0.718) & & (0.547--0.720) \\
\bottomrule
\end{tabular}
\end{table}

\begin{figure}[t]
\centering
\includegraphics[width=\columnwidth]{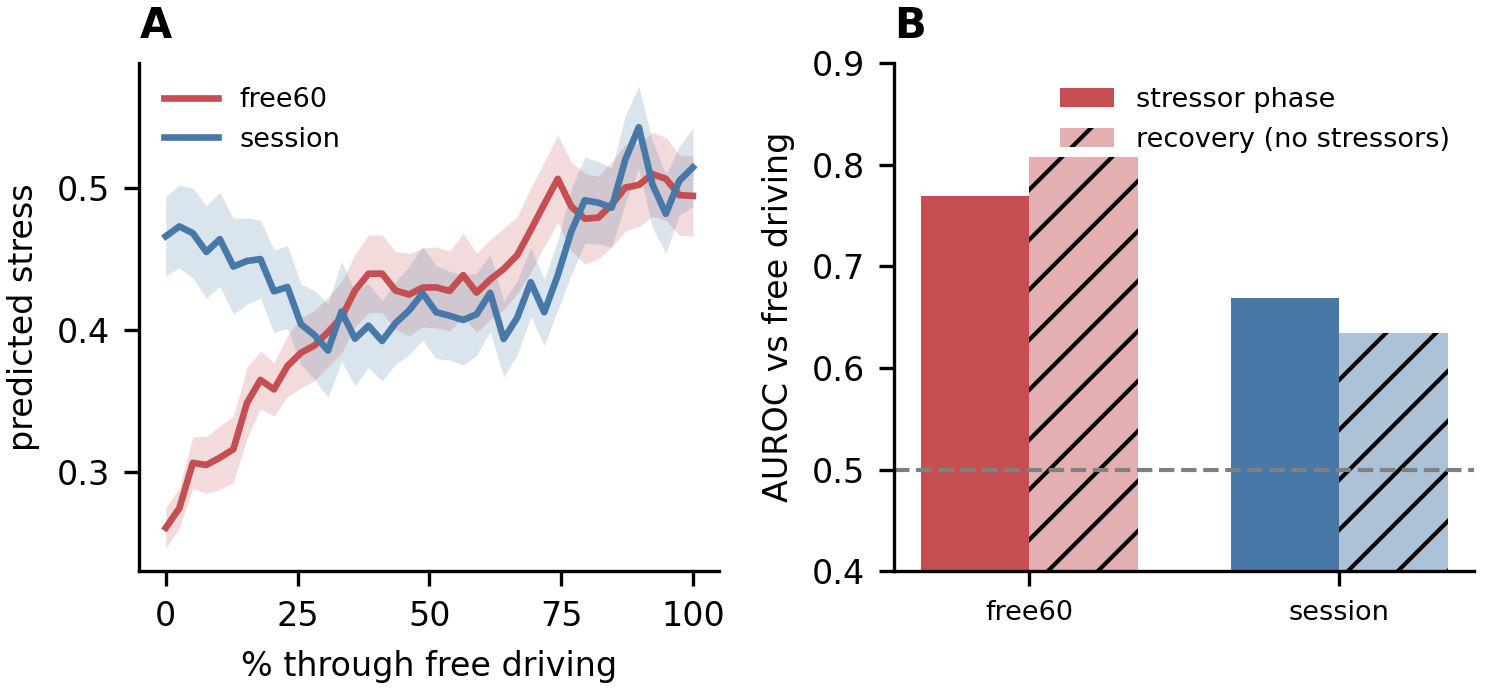}
\vspace{-2mm}
\caption{Predicted stress tracks session time under the \texttt{free60} baseline. \textbf{(A)} Averaged predicted stress across the free-driving phase, in which any upward trend is attributable to elapsed time rather than stressor manipulation. Shaded bands are standard errors across sessions. \textbf{(B)} Discrimination of the stressor phase (solid) and of the recovery phase (hatched) from the free-driving phase. With the \texttt{free60} reference the recovery phase is predicted higher than the stressor phase, whereas the \texttt{session} reference reflects the expected stress regulation effect. The dashed line marks chance.}
\label{fig:drift}
\vspace{-3mm}
\end{figure}

\subsection{Drift-aware Model Performance}

We first evaluate our stress detection classifier on the primary task of distinguishing between the free-driving phase and the subsequent stressor phase. Results for the multimodal setting (23 features) are reported in Table~\ref{tab:baselines} for each of the three baseline normalization methods. Normalization using the initial free-driving interval yields the best phase discrimination with an average AUROC of 0.768. In contrast, whole-session normalization substantially outperforms the other two methods on the more stringent within-phase metric, which evaluates the ability to track induced stressors within the same session. This discrepancy is explained by the \textit{Drift} metric, which indicates that normalization based on the initial free-driving interval is confounded by sensor drift over time, effectively causing the model to distinguish temporally adjacent periods rather than stress states. As further evidence of this confound, the recovery period, which is stressor-free by design, is not only predicted as stressful under this scheme but is predicted as more stressful than the stressor phase itself. Finally, we note that an external baseline derived from participants' practice sessions did not manage to form a performant baseline for this task.

\begin{table}[t]
\centering
\caption{Performance of the stressor-locked classifier by modality, with 95\% bootstrap confidence intervals over participants. The classifier is trained on the \textit{within-phase} contrast: $P_\text{control}$ and $P_\text{post-onset}$ are the mean probabilities assigned to the two window classes, both drawn from the stressor phase. Wilcoxon $p$ is one-sided signed-rank against chance.}

\label{tab:clf_scores}
\small
\setlength{\tabcolsep}{5.5pt}
\begin{tabular}{@{}lcccc@{}}
\toprule
Features & $P_\text{control}$ & $P_\text{post-onset}$ & AUROC & Wilcoxon \\
 & & & within-phase & $p$ \\
 & (95\% CI) & (95\% CI) & (95\% CI) & \\
\midrule
ECG, PPG & 0.495 & 0.520 & 0.576 & 0.017 \\
 & (0.481--0.508) & (0.508--0.531) & (0.527--0.623) & \\
\addlinespace
EDA & 0.480 & 0.539 & 0.624 & 0.014 \\
 & (0.454--0.508) & (0.515--0.559) & (0.555--0.682) & \\
\addlinespace
RSP & 0.500 & 0.507 & 0.528 & 0.197 \\
 & (0.489--0.511) & (0.498--0.516) & (0.484--0.571) & \\
\addlinespace
SKT & 0.479 & 0.546 & 0.651 & \textless 0.001 \\
 & (0.447--0.511) & (0.526--0.564) & (0.584--0.712) & \\
\addlinespace
All & 0.483 & 0.567 & \textbf{0.661} & \textless 0.001 \\
 & (0.454--0.512) & (0.540--0.592) & (0.594--0.718) & \\
\bottomrule
\end{tabular}
\vspace{-0.2cm}
\end{table}

As a result, in the following analyses, we use \texttt{session} normalization for model training, evaluation, and analysis. Table~\ref{tab:clf_scores} reports the full classification results for the multimodal model alongside modality-specific results. The four physiological modalities perform comparably on the drift-controlled contrast. Skin temperature and electrodermal activity emerge as the primary drivers of the detected stress responses during driving, whereas respiration-related features are the least discriminating, achieving chance-level performance in the within-phase evaluation. Multimodal fusion yields the best overall performance; however, it does not significantly outperform skin temperature alone, particularly when distinguishing between individual stressors.

\subsection{Stressor-wise Discrimination}

The within-phase performance metric allows us to assess whether the model is sensitive to genuine stressor-evoked responses rather than temporal sensor drift. This metric is de-confounded by contrasting windows occurring entirely within 30 seconds of stressor onset with windows occurring more than 30 seconds from any onset, while sampling both classes from the same stressor phase. We further verified that performance was insensitive to the choice of control-window distance across a range of 15–60 seconds ($p=0.681$). We therefore use this metric to evaluate the efficacy of each individual stressor in the study.

\begin{figure}[t]
\centering
\includegraphics[width=\columnwidth]{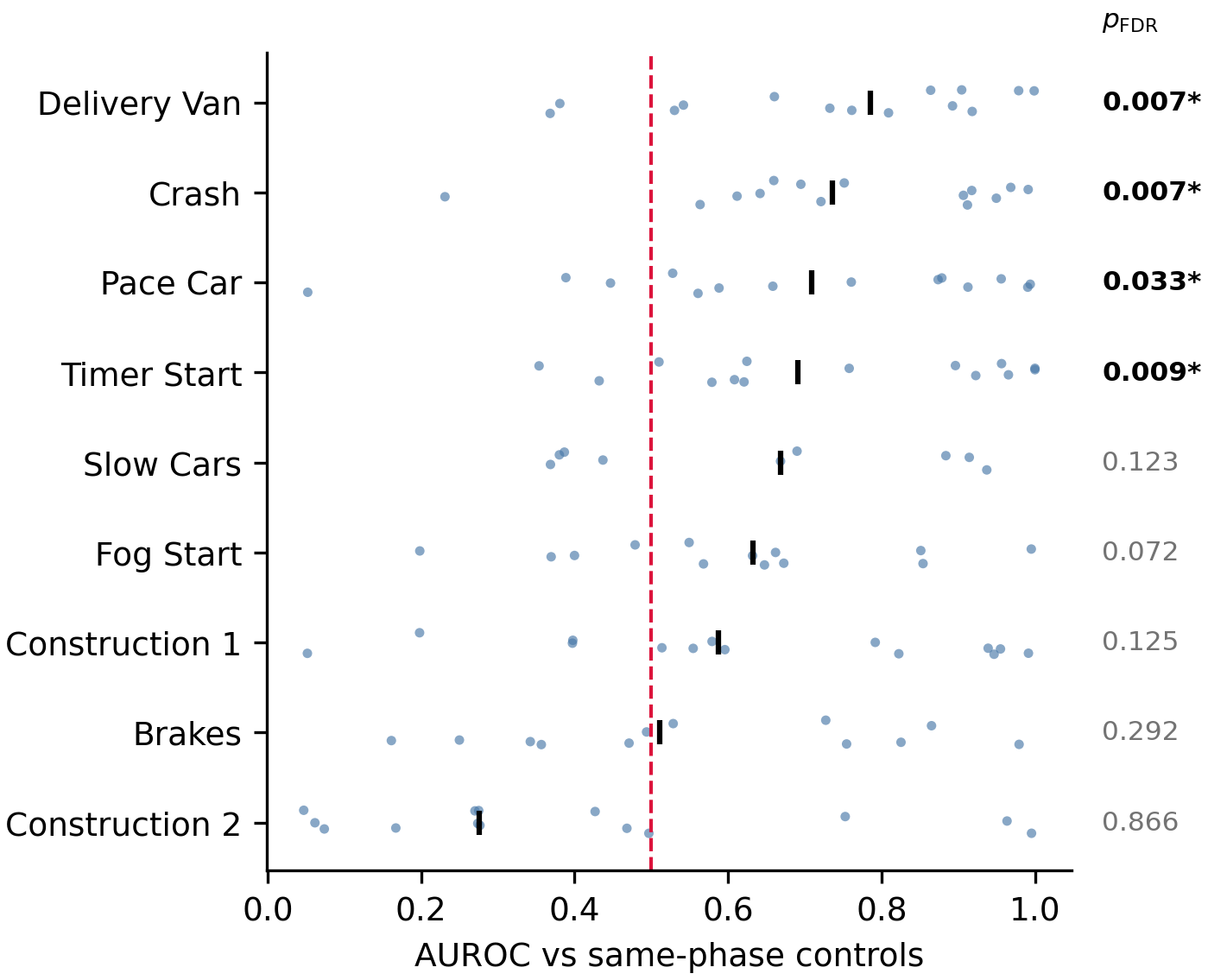}
\vspace{-2mm}
\caption{Stress estimation around each induced stressor, measured against same-phase controls. Each point is one session and the vertical bar is the median and stressors are ordered by median response. The dashed line marks random chance. Values in the right-hand column are one-sided Wilcoxon signed-rank tests of the participant scores against chance, FDR-corrected. Those reaching $p_{\mathrm{FDR}} < 0.05$ are bolded. Note: The \textit{Barrel} event was omitted as it happened at the end of the session.}
\label{fig:perstressor}
\vspace{-3mm}
\end{figure}

In Figure~\ref{fig:perstressor} we report the within-phase discriminability of each stressor category in terms of AUROC, together with one-sided Wilcoxon signed-rank tests against chance. FDR-corrected $p$-values indicate significant effects for several events, including the delivery van (0.785), the crash (0.736), the pace car (0.710) and the timer (0.691). Because several stressors occurred multiple times within a session, responses to repeated instances were averaged within each session before statistical testing. Stressors that impose a sustained constraint on the driver, such as the delivery van, the pace car and the timer, produced detectable responses, as did the single high-intensity surprise event, whereas the recurring low-intensity events, such as sudden braking and fog, did not. The response to the second construction zone was also markedly weaker than that to the first, consistent with the effect of habituation. Specifically, stress predictions were generally higher during the first construction zone by a median difference of 0.127, with this pattern observed in 12 of 14 sessions ($p = 0.0004$, one-sided paired test). Because this effect runs opposite to temporal drift, it could have been obscured under normalization procedures that do not control for temporal drift, as identified in Table~\ref{tab:baselines}.

\subsection{Feature Attribution}

\begin{figure}[t]
\centering
\includegraphics[width=\columnwidth]{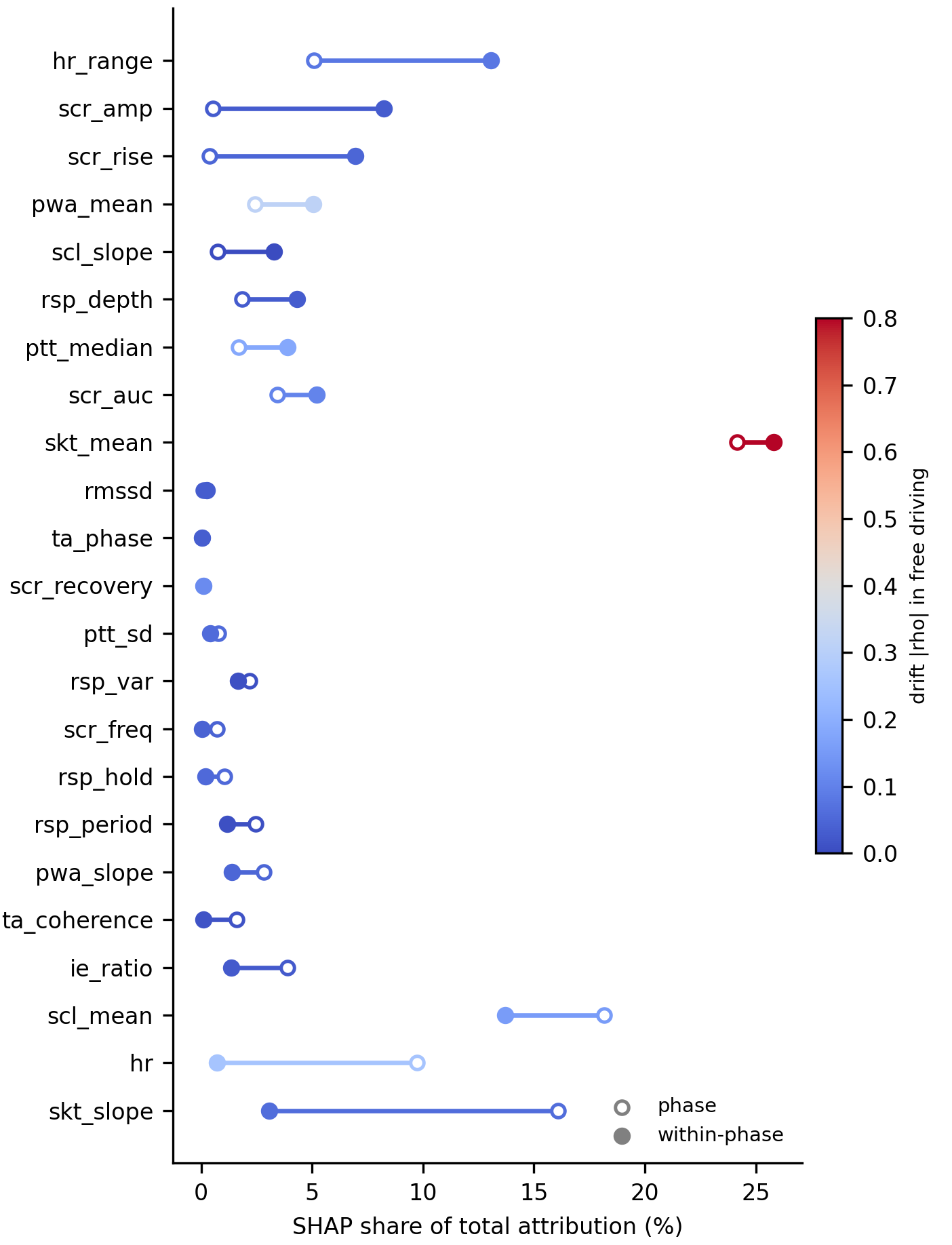}
\caption{Mean absolute SHAP attribution for each feature as a percentage of total attribution, for models trained using the phase (open markers) and the within-phase contrast (filled markers). Features are ordered by the magnitude of the change in attribution between models. Color indicates temporal drift, quantified for each feature as the absolute Spearman correlation with time during free driving. Under the drift-controlled within-phase contrast, attribution shifts from slowly varying level features toward features capturing event-scale responses.}
\label{fig:shap}
\vspace{-2mm}
\end{figure}

We next examined whether this dissociation between models relying on generic drift patterns and those detecting genuine stressor effects led the model to rely on different physiological features. We trained the same multimodal model separately on the phase and within-phase contrasts and computed SHAP~\cite{lundberg2017unified} attributions on a common set of held-out participants. Under \texttt{session} normalization, the combined attribution of the three phasic skin-conductance features and heart-rate range increased from 9.4\% to 33.5\%. On the other hand, the combined attribution of skin-temperature slope, heart rate, and tonic skin-conductance level decreased from 44\% to 17.5\%. Each individual change was significant at the participant level ($p<10^{-4}$). Thus, removing cumulative temporal variation shifted the model more toward transient autonomic responses than tonic levels.

Mean skin temperature remained the most influential individual feature under both normalization schemes, despite also being the feature exhibiting the strongest temporal drift. More generally, however, the magnitude of drift across the 23 features did not predict the corresponding change in attribution ($\rho=0.04$, $p=0.86$). High attribution should therefore not be interpreted as evidence that the model uses a feature as a monotonic marker of stress. Achieving an AUROC of 0.661 instead suggests that skin temperature was likely used as a conditioning variable within the depth-two interactions available to the model.

\begin{table*}[t]
\centering
\caption{Association between driving behavior and stress across four analyses, with 95\% confidence intervals in parentheses. \emph{Near vs far} contrasts the telemetry itself between post-onset and control windows, tested with a Wilcoxon signed-rank over participants. \emph{All phases} and \emph{stressor phase} are within-participant coefficients from mixed models relating the stressor-locked index to each vehicle feature, the first fitted across the whole recording, the second restricted to stressor-phase windows. In both, the predictor is centered within participant and phase exactly as the driving signal is standardized, and the position within the phase is held constant. \emph{Per event} correlates the size of the physiological response at each onset with the size of the accompanying change in driving. Marked entries are significant after FDR correction across the 10 vehicle features within each analysis.}
\label{tab:behavior}
\setlength{\tabcolsep}{8pt}
\footnotesize
\begin{tabular}{@{}lcccc@{}}
\toprule
 & \textbf{Near vs far} & \textbf{All phases} & \textbf{Stressor phase} & \textbf{Per event} \\
\textbf{Feature} & AUROC & Coefficient $\beta$ & Coefficient $\beta$ & Correlation $\rho$ \\
\midrule
\quad Steering SD        & \(0.625^{\ast}\,(0.559,0.705)\,\) & \(+0.162^{\ast}\,(+0.103,+0.222)\,\) & \(+0.124^{\ast}\,(+0.046,+0.202)\,\) & \(+0.417^{\ast}\,(+0.236,+0.595)\,\) \\
\quad Steering entropy   & \(0.484\phantom{^{\ast}}\,(0.439,0.572)\,\) & \(-0.087^{\ast}\,(-0.147,-0.027)\,\) & \(-0.044\phantom{^{\ast}}\,(-0.126,+0.037)\,\) & \(-0.127\phantom{^{\ast}}\,(-0.286,-0.053)\,\) \\
\specialrule{\lightrulewidth}{0.15em}{0.15em}
\quad Speed SD (eng)     & \(0.689\phantom{^{\ast}}\,(0.604,0.783)\,\) & \(+0.131^{\ast}\,(+0.071,+0.191)\,\) & \(+0.193^{\ast}\,(+0.112,+0.274)\,\) & \(+0.236\phantom{^{\ast}}\,(+0.086,+0.418)\,\) \\
\quad Speed SD           & \(0.580\phantom{^{\ast}}\,(0.500,0.743)\,\) & \(+0.131^{\ast}\,(+0.072,+0.190)\,\) & \(+0.169^{\ast}\,(+0.088,+0.250)\,\) & \(-0.045\phantom{^{\ast}}\,(-0.164,+0.312)\,\) \\
\specialrule{\lightrulewidth}{0.15em}{0.15em}
\quad Throttle magnitude & \(0.666\phantom{^{\ast}}\,(0.586,0.780)\,\) & \(+0.089^{\ast}\,(+0.030,+0.149)\,\) & \(+0.131^{\ast}\,(+0.050,+0.211)\,\) & \(+0.104\phantom{^{\ast}}\,(-0.244,+0.314)\,\) \\
\quad Throttle entropy   & \(0.610^{\ast}\,(0.573,0.727)\,\) & \(+0.026\phantom{^{\ast}}\,(-0.033,+0.085)\,\) & \(-0.027\phantom{^{\ast}}\,(-0.106,+0.052)\,\) & \(+0.221\phantom{^{\ast}}\,(+0.212,+0.464)\,\) \\
\quad Throttle rate      & \(0.533\phantom{^{\ast}}\,(0.481,0.737)\,\) & \(+0.017\phantom{^{\ast}}\,(-0.042,+0.076)\,\) & \(+0.028\phantom{^{\ast}}\,(-0.053,+0.109)\,\) & \(+0.066\phantom{^{\ast}}\,(-0.168,+0.227)\,\) \\
\specialrule{\lightrulewidth}{0.15em}{0.15em}
\quad Brake magnitude    & \(0.515\phantom{^{\ast}}\,(0.444,0.583)\,\) & \(+0.062\phantom{^{\ast}}\,(+0.004,+0.120)\,\) & \(+0.110^{\ast}\,(+0.029,+0.192)\,\) & \(+0.228\phantom{^{\ast}}\,(+0.007,+0.382)\,\) \\
\quad Brake entropy      & \(0.453\phantom{^{\ast}}\,(0.338,0.509)\,\) & \(-0.001\phantom{^{\ast}}\,(-0.060,+0.058)\,\) & \(+0.000\phantom{^{\ast}}\,(-0.079,+0.080)\,\) & \(-0.125\phantom{^{\ast}}\,(-0.682,+0.006)\,\) \\
\quad Brake rate         & \(0.446\phantom{^{\ast}}\,(0.250,0.520)\,\) & \(-0.003\phantom{^{\ast}}\,(-0.062,+0.055)\,\) & \(-0.023\phantom{^{\ast}}\,(-0.102,+0.056)\,\) & \(-0.202\phantom{^{\ast}}\,(-0.310,+0.068)\,\) \\
\bottomrule
\end{tabular}
\vspace{-3mm}
\end{table*}

Indeed, univariate comparisons between post-onset and control windows provided a complementary picture. Two features differed significantly between the two conditions: heart-rate range increased following stressor onset ($+0.332$, $p = 0.0001$), whereas respiratory period decreased ($-0.200$, $p = 0.045$). Skin-conductance response area and amplitude shifted in the expected direction too. By contrast, mean skin temperature and tonic skin conductance received high model attribution but showed no consistent univariate direction. We therefore interpret them as contextual features used through nonlinear interactions, rather than as variables that consistently increase or decrease in response to stress.
\vspace{-2mm}

\subsection{Correlates of Driving Behavior}
\label{sec:behavior}

Finally, we examined whether our stressor-locked classifier corresponded to observable, generalizable changes in driving, as measured from the 7\,Hz vehicle telemetry. First, applying the \textit{within-phase} comparison directly to the telemetry showed that 2 of the 10 vehicle features differed significantly between post-onset and control windows across participants, namely steering variability and throttle usage entropy. The remaining features shifted in the expected direction but did not survive correction at the participant level.

We also tested whether variation in the model-predicted physiological response modulated driving behavior. For each telemetry feature, we fitted a mixed-effects model relating the feature to the model output, with a participant-specific random intercept. The predictor was centered within participant and phase, and position within phase was included as a covariate. This formulation removes free-driving versus stressor-phase differences for each participant and controls for gradual changes within phase. Across the full recording, higher predicted stress was associated with greater steering variability, speed variability, throttle magnitude, and lower steering entropy. When the analysis was restricted to stressor-phase windows, associations remained for steering variability, speed variability, and throttle magnitude, while brake magnitude also emerged as a significant correlate. By contrast, the phase-contrast index retained no significant associations under either specification. Notably, despite its higher nominal discrimination performance, the phase-contrast index did not explain variation in driving behavior once phase and position within phase were controlled.

Last, we asked whether the magnitude of the physiological response to individual stressors covaried with the magnitude of the accompanying behavioral change. Across the 16 participants with sufficient events, larger physiological responses were associated with greater changes in steering variability, with engine speed variability close to threshold. As a result, events that elicited stronger physiological responses also produced larger changes in vehicle control. Together, these results suggest that a physiological stressor estimator is capable of capturing behaviorally relevant variation in the stability and intensity of vehicle control.\vspace{-3mm}

\section{Conclusion}

In this study we developed a robust feature-based estimator of driving stress and evaluated its sensitivity to prespecified stressor events while controlling for temporal drift. The resulting model distinguished post-onset of stressors from temporally matched control windows with an AUROC of 0.661. Under this modeling assumption, sustained and intense stressors produced more consistent responses and model attribution shifted from slow tonic changes toward phasic autonomic biomarkers. The resulting model was also reflected in participants' driving behavior, since events that elicited stronger physiological responses were accompanied by greater changes in steering, speed, and pedal control.

These findings support event-locked modeling and temporally controlled evaluation as important foundations for reliable physiological monitoring of driving stressors. Several limitations should nevertheless be acknowledged. The study used a modest sample size and a controlled simulator that cannot reproduce the consequences or full arousal range of naturalistic driving. Moreover, the labels denote exposure to designed stressors rather than directly reported stress, and the detected responses may therefore reflect broader autonomic arousal or task demand. Validation using randomized event schedules, naturalistic driving, larger samples, and complementary contextual annotations is therefore needed to establish the generalizability of the stressor-locked response.\vspace{-2mm}

\section*{Acknowledgments}

This study was supported by Toyota Motor Engineering \& Manufacturing North America and MIRISE Technologies. 

\section*{Data \& Code Availability}

The code used to conduct the experiments and the analysis in this study will be publicly available soon on the following GitHub link: https://github.com/usc-sail/tracking-stressors. The dataset presented and used in this study can be accessed via IEEE DataPort: https://dx.doi.org/10.21227/x51t-jr44
\vspace{-2mm}

\section*{Competing Interests}

K.A., F.M.Q.S., E.M.D., and P.S. have a US patent pending 18817888 titled ``Systems and Methods for Vehicle Biometrics Processing and Vehicle Control." All other authors declare no competing interests.

\ifreferencesused
    \bibliographystyle{IEEEtran}
    \bibliography{refs}
\fi

\end{document}